\documentclass[letterpaper, 10 pt, conference]{IEEEconf}  

\IEEEoverridecommandlockouts                              
\usepackage{amssymb}
\usepackage{graphics}
\usepackage{epsfig}
\usepackage{mathptmx}
\usepackage{times}
\usepackage{amsmath}
\usepackage{footmisc}

\newtheorem{theorem}{Theorem}

\newtheorem{definition}{Definition}
\newtheorem{example}{Example}

\title{\LARGE \bf Hybrid Systems and Control With Fractional Dynamics (I): Modeling and Analysis}
\author{S. Hassan~HosseinNia, In\'{e}s~Tejado, and~Blas~M.~Vinagre
\thanks{This work was supported by the Spanish Ministry of Economy and Competitiveness under 
the research project DPI2012-37062-C02-02.}
\thanks{S. H. HosseinNia, I. Tejado and B. M. Vinagre are with Department of Electrical, Electronic and Automation Engineering, Industrial Engineering School, University of Extremadura, 06006 Badajoz, Spain. e-mail: \tt \{hoseinnia,itejbal,bvinagre\}@unex.es}}

\begin{document}

\maketitle
\thispagestyle{empty}
\pagestyle{empty}

\begin{abstract}
No mixed research of hybrid and fractional-order systems into a cohesive and multifaceted whole can be found in the literature. This paper focuses on such a synergistic approach of the theories of both branches, which is believed to give additional flexibility and help to the system designer. It is part I of two companion papers and introduces the fundamentals of fractional-order hybrid systems, in particular, modeling and stability analysis of two kinds of such systems, i.e., fractional-order switching and reset control systems. Some examples are given to illustrate the applicability and effectiveness of the developed theory. Part II will focus on fractional-order hybrid control.
\end{abstract}

\section{Introduction}
Hybrid systems (HS) are heterogeneous dynamic systems whose behaviour is determined by interacting continuous-variable and discrete-event dynamics, and they arise from the use of finite-state logic to govern continuous physical processes or from topological and networks constraints interacting with continuous control \cite{Gollu_89,schumacher_99,Goebel_09}. It is worth mentioning that, among them, we focuses on two kinds of HS in this work: switching and reset control systems. Switching systems, a class of HS consisting of several subsystems and a switching rule indicating the active subsystem at each instant of time, have been the subject of interest for the past decades, for their wide application areas. Likewise, reset control systems are standard control systems endowed with a reset mechanism, i.e., a strategy that resets to zero the controller state (or part of it) when some condition holds. The hybrid behaviour comes from the instantaneous jump due to resets of whole or part of system states \cite{banos2011,Schutter2009}.

Many real dynamic systems are better characterized using a fractional-order dynamic model based on differentiation and integration of non-integer-order. The concept of fractional calculus has tremendous potential to change the way we see, model, and control the nature around us. Denying fractional derivatives is like saying that zero, fractional, or irrational numbers do not exist. From the control engineering point of view, improving and developing the control is the major concern (see e.g. \cite{Monje10,Podlubny_99a}). 

Recently, the wide applicability of both HS and systems with  fractional-order dynamics has inspired a great deal of research and interest in both fields. Unfortunately, in general there are many difficulties in mixing different mathematical domains. The case of combining the theories of such systems is no exception. Given this motivation, this paper arises from the idea of coupling two different distinct branches of research, fractional calculus and HS, into a synergistic way, which is believed to give additional flexibility and help to the system designer, taking advantage of the potentialities of both worlds. To this respect, a mathematical framework of fractional-order hybrid systems (FHS), including modeling, stability analysis, control and simulation, is required to be developed. Accordingly, the objective of part I of these two companion papers is to introduce the mentioned framework of HS with fractional-order dynamics, namely, modeling and analysis issues. 

The remainder of part I of this paper is organized as follows. In Section \ref{model}, modeling of FHS is presented through differential inclusions (DI); two special examples of switching and reset control systems are studied. Section \ref{SFHSF} addresses stability analysis of such systems. Three stability examples, again for switching and reset control systems, are given to show the applicability of the developed theory. The concluding remarks are drawn in Section \ref{Concf}.

\section{Modeling of Fractional-Order Hybrid Systems}
\label{model}

This section deals with fundamentals of two kinds of fractional-order hybrid systems, i.e., switching systems and reset control systems based on fractional-order differential inclusions (FDI). Then, two special HS are modeled.

\subsection{Fractional-order differential inclusions}

A widely used model of a continuous-time dynamical system is the first-order
differential equation $\dot{x}=f(x,u)$, with $x$ and $u$ belonging to an $n$-dimensional 
Euclidean space $\mathbb{R}^{n}$. This model can be expanded in two directions that are relevant for
HS. First, we can consider differential equations with state
constraints, that is, $\dot{x}=f(x,u)$ and $x\in C,\ u\in C_{u}$, where flow sets $C$
and $C_{u}$ are subsets of $\mathbb{R}^{n}.$ Second, we can consider the situation where the right-hand side of
the DI is replaced by a set that may depend on $x$. Both
situations lead to the DI $\dot{x}\in F(x)$, where $F$
is a set-valued mapping. Likewise, the combination of the two generalizations leads to
constrained DI as follows: $\dot{x}\in F(x,u),\ x\in C,\ u\in C_{u}$.

A typical model of a discrete-time dynamical system is the first-order
equation $x^{+}=g(x,u)$, with $x,u\in 
\mathbb{R}
^{n}$. The notation $x^{+}$ indicates that the next value of the state is
given as a function of the current state $x$ through the value $g(x)$. As
for differential equations, it is a natural extension to consider
constrained difference equations and difference inclusions, which leads to
the model $x^{+}\in G(x,u),\ x\in D,\ u\in D_{u}$, where $G$ is a set-valued
mapping and jump sets $D$ and $D_{u}$ are subsets of $\mathbb{R}^{n}.$ Since a model of a hybrid 
dynamical system requires a description of the time driven dynamics, the event driven 
dynamics, and the regions on which these dynamics apply, we include both a constrained 
DI and a constrained difference inclusion in a general model of a HS in the form
\begin{equation}
\begin{matrix}
\dot{x}\in F(x,u), x \in C, u\in C_{u}, \\ 
x^{+}\in G(x,u), x \in D, u\in D_{u}.
\end{matrix}
\label{Diff_Inc}
\end{equation}
Taking into account integer-order DI described by (\ref{Diff_Inc}), its generalization to fractional-order can be expressed as
\begin{eqnarray}
\label{FDiff_Inc}
D^{\alpha }x\in F(x,u), x\in C, u\in C_{u}, \\ \nonumber
x^{+}\in G(x,u), x\in D, u\in D_{u},
\end{eqnarray}
where $D^{\alpha}$ is the fractional-order operator with $\alpha\in\mathbb{R}$.

\subsection{Switching systems}
Switching system is a hybrid dynamical system consisting of a family of continuous-time subsystems
and a rule that orchestrates the switching among them \cite{Sun_05}. A general formulation of the switching systems with fractional-order is:
\begin{equation}
D^{\alpha}{x}=Ax, A \in co\left\{ A_{1}, ..., A_{L} \right\}.
\label{FSWHM}
\end{equation}
where $co$ denotes the convex combination and $A_{i}$, $i=1,...,L$, is the switching subsystem. A primary motivation for studying such systems came partly from the fact that switching systems and switching multi-controller systems have numerous applications in control of mechanical systems, process control,
automotive industry, power systems, traffic control, and so on. Let us now model switching system of multi-controller by means of FDI in the following example.

\begin{example} Modelling of a fractional-order multi-controller system
\label{Exmulti}
\end{example}

\begin{figure}[htbp]
\begin{center}
\includegraphics[width=0.36\textwidth]{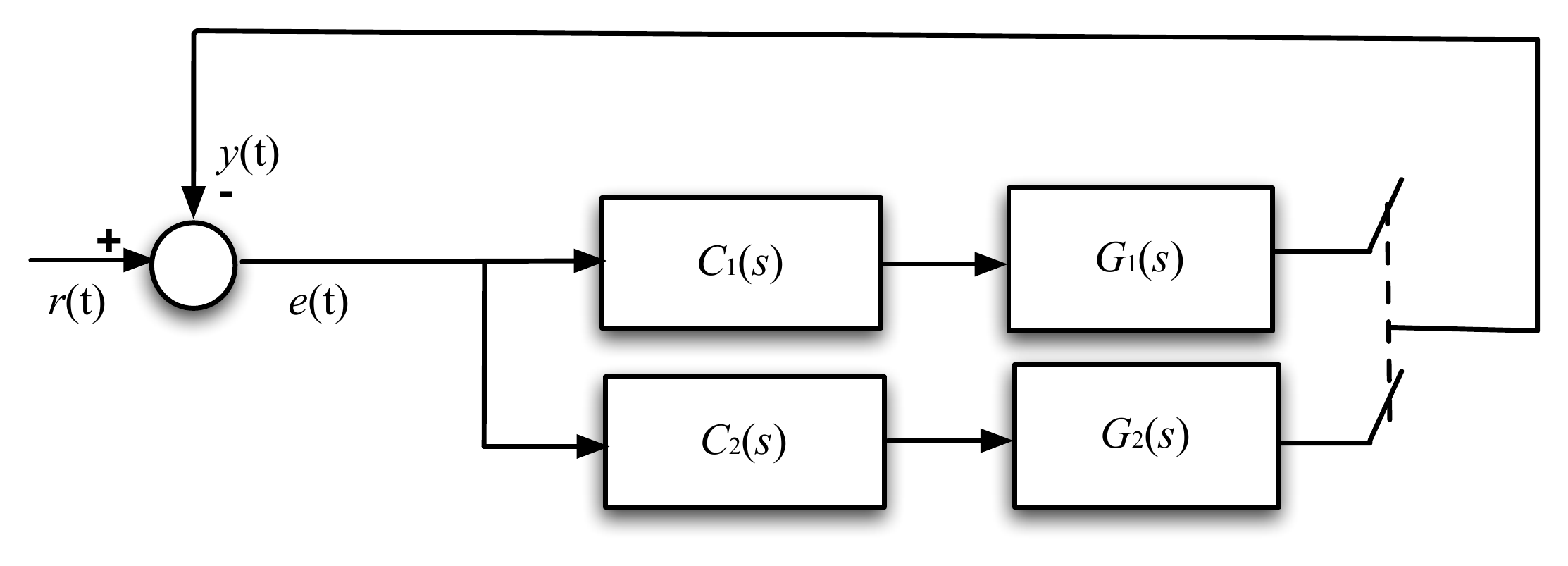}
\caption{Closed-loop system with two controllers}
\label{Multi_Frac}
\end{center}
\end{figure}

Let us consider a first-order system with two different dynamics as follows (see Fig. \ref{Multi_Frac}):
\begin{equation}
G_i(s)=\frac{K_i}{s+\tau_i}, i=\{1, 2\}, 
\end{equation}
controlled by the following fractional-order PI controllers:
\begin{equation}
C_i(s)=k_{p_i}+\frac{k_{i_i}}{s^{\alpha_i}}, i=\{1, 2\}.
\end{equation}
Then, the closed-loop transfer function of the system can be written as:
\begin{equation}
\label{TFM}
\frac{Y(s)}{R(s)}=\frac{a_is^{\alpha_i}+b_i}{s^{\alpha_i+1}+(\tau_i+a_i)s^{\alpha_i}+b_i}, i=\{1, 2\},
\end{equation}
where $a_i=K_ik_{p_i}$ and $b_i=K_ik_{i_i}$. Assuming $\alpha_i=\frac{q_i}{p_i}$, the state space form of (\ref{TFM}) is given by:
\tiny
\begin{equation}
\begin{bmatrix}
D^{\frac{1}{q_i}}x_1\\ 
D^{\frac{1}{q_i}}x_2\\ 
\vdots \\ 
D^{\frac{1}{q_i}}x_{p_i+1}\\ 
\vdots\\
D^{\frac{1}{q_i}}x_{p_i+q_i-1}\\
D^{\frac{1}{q_i}}x_{p_i+q_i}
\end{bmatrix}=\begin{bmatrix}
0 & 1 & 0 &\cdots   & 0 & \cdots & 0\\ 
 0& 0 & 1 & \cdots &0 & \cdots & 0 \\ 
\vdots  & \vdots & \vdots & \ddots  & \vdots & \ddots & \vdots\\ 
0 & 0 & 0 & \cdots  & 1 & \cdots & 0\\ 
\vdots  & \vdots & \vdots & \ddots  & \vdots & \ddots & \vdots\\ 
0 & 0 & 0 & \cdots & 0 & \cdots & 1\\
 -b_i & 0 & 0  & \cdots & -(\tau_i+a_i) & \cdots & 0
\end{bmatrix}\begin{bmatrix}
x_1\\ 
x_2\\ 
\vdots \\ 
x_{p_i+1}\\ 
\vdots \\
x_{p_i+q_i-1}\\
x_{p_i+q_i-1}
\end{bmatrix}+
\begin{bmatrix}
0\\ 
0\\ 
0\\ 
\vdots \\ 
0\\ 
\vdots \\ 
1
\end{bmatrix}U(r(t)), i=\{1, 2\},
\end{equation}
\normalsize
where $U(r(t))=a_iD^{\alpha_i}r(t)+b_ir(t)$. It is obvious that the closed-loop system can be written in a general form as:
\begin{equation}
D^{\alpha_i} x=A_ix+B_iU_i.
\end{equation}
Now assume that the controller one $C_1(e)$ will be activated if $e=r(t)-y(t)>-\varepsilon$, whereas the controller $C_2(e)$ will be activated if $e=r(t)-y(t)<\varepsilon$. Thus, the FDI are taken to be:
\begin{equation}
\begin{bmatrix}
D^{\alpha_i} x \\
D^{\alpha_i} i 
\end{bmatrix}
=\begin{bmatrix}
A_ix+B_iU_i \\
0
\end{bmatrix},
\end{equation}
The flow set and jump set are respectively taken as:
\begin{eqnarray}
\nonumber C:=\left \{ (x,i) \in \mathbb{R}^{\alpha_i+1} \times \left \{1, 2 \right \} | i=1\& y(t)<r(t)+\varepsilon \text{ or } \right. \\ \left.  i=2  \& y(t)>r(t)-\varepsilon \right \},
\end{eqnarray}
and
\begin{eqnarray}
\nonumber D:=\left \{ (x,i) \in \mathbb{R}^{\alpha_i+1} \times \left \{1, 2 \right \} | i=1 \& y(t)=r(t)+\varepsilon \text{ or }  \right. \\ \left. i=2 \& y(t)=r(t)-\varepsilon \right \}.
\end{eqnarray}

In what concerns the jump map, since the role of jump changes is to toggle the logic mode and the state component $x$ does not change during jumps, the jump map will be:
\begin{equation}
\begin{bmatrix}
x \\
i
\end{bmatrix}^+
=\begin{bmatrix}
x \\
3-i
\end{bmatrix}.
\end{equation}

\subsection{Fractional-order reset control systems}
\label{Dreset}



\begin{figure}[ptbh]
\begin{center}
\includegraphics[width=0.36\textwidth]{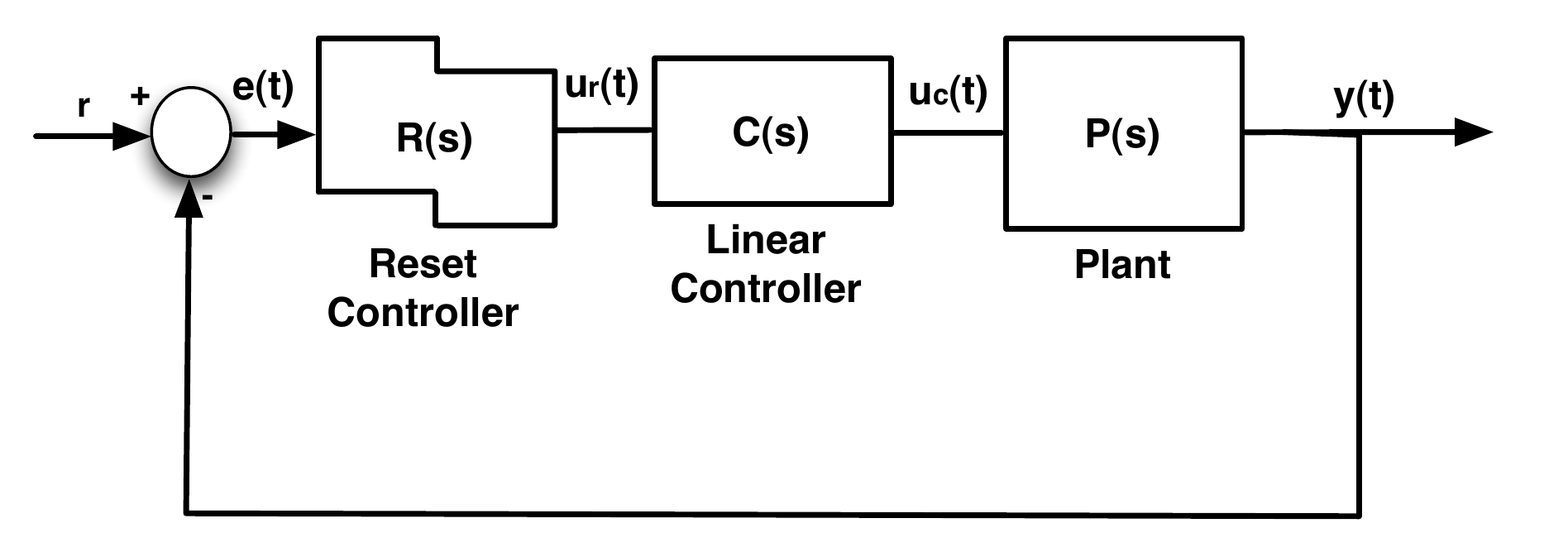}
\end{center}
\caption{Block diagram of a reset control system}
\label{Reset_Block}%
\end{figure}

Let us now model reset control systems by means of FDI. The block diagram of a general reset control system is shown in Fig. \ref{Reset_Block}. It can be observed that the dynamics of the reset controller can be described by the FDI equation as:
\begin{eqnarray}
\begin{matrix}
D^\alpha x_r(t)=A_r x_r(t)+B_re(t), \ e(t)\neq0,\\ 
x_r(t^+)=A_{R_r}x_r(t), \ e(t)=0,\\
u_r(t)=C_rx_r(t)+D_re(t),
\end{matrix}
\label{reseteq}
\end{eqnarray}
where $0< \alpha \leq 1$ is the order of differentiation, $x_r(t) \in \mathbb{R}^{n_r}$ is the reset controller state vector and $u_r(t) \in \mathbb{R}$ is its output.
The matrix $A_{R_r}  \in \mathbb{R}^{n_r\times n_r}$ identifies that subset of states $x_r$ that are reset (the last ${\mathcal{R}}$ states) and use the structure $A_{R_r}=\begin{bmatrix}
I_{n_{\bar{\mathcal{R}}}}& 0 \\ 
0 & 0_{n_{\mathcal{R}}}
\end{bmatrix}$ and $n_{\bar{\mathcal{R}}}=n_r-n_{\mathcal{R}}$.

The linear controller $C(s)$ and plant $P(s)$ have, respectively, state space representations as follows:
\begin{eqnarray}
\begin{matrix}
D^\alpha x_c(t)=A_cx_c(t)+B_cu_r(t),\\ 
u_c(t)=C_cx_c(t),
\end{matrix}
\label{Conteq}
\end{eqnarray}
and
\begin{eqnarray}
\begin{matrix}
D^\alpha x_p(t)=A_px_p(t)+B_pu_c(t),\\ 
y(t)=C_px_p(t),
\end{matrix}
\label{Syseq}
\end{eqnarray}
where $A_p\in \mathbb{R}^{n_p \times n_p}$, $B_p \in \mathbb{R}^{n_p\times 1}$, $C_p \in \mathbb{R}^{1\times n_p}$, $A_c\in \mathbb{R}^{n_c\times n_c}$, $B_c \in \mathbb{R}^{n_c\times 1}$ and $C_c \in \mathbb{R}^{1\times n_c}$. 

The closed-loop reset control system can be then described by the following FDI:
\begin{eqnarray}
\begin{matrix}
D^\alpha x(t)=A_{cl}x(t)+B_{cl}r, \ x(t) \notin  \mathcal{M}\\ 
x(t^+)=A_{R}x(t), \  x(t)\in \mathcal{M}\\
y(t)=C_{cl}x(t)
\end{matrix}
\label{CLeq}
\end{eqnarray}
where
$x=\begin{bmatrix}
x_p\\ 
x_c\\ 
x_r
\end{bmatrix}$,
$A_{cl}=\begin{bmatrix}
 A_p& B_pC_c & 0 \\ 
 -B_cD_r  C_p & A_c & B_cC_r \\ 
 -B_rC_p& 0 & A_r
\end{bmatrix}$,
$A_{R}=\begin{bmatrix}
 I_{n_p} & 0 & 0 \\ 
 0 & I_{n_c}& 0 \\ 
 0 & 0 & A_{R_r}
\end{bmatrix}$, $B_{cl}=\begin{bmatrix}
0 & B_cD_r & B_r\end{bmatrix}^T$ and $C_{cl}=\begin{bmatrix}
C_p & 0 & 0 \end{bmatrix}$. The reset surface $\mathcal{M}$ is defined by:
\begin{eqnarray}
\begin{matrix}
\mathcal{M}=\left \{ x\in\mathbb{R}^n: C_{cl}x=r,\ (I-A_R)x\neq0 \right \}.
\end{matrix}
\label{surf}
\end{eqnarray}
where $n=n_r+n_c+n_p$.
In absence of the linear controller $C(s)$, the state space realization of the closed-loop system can be also stated as (\ref{CLeq}) with
$x=\begin{bmatrix}
x_p\\ 
x_r
\end{bmatrix}$,
$A_{cl}=\begin{bmatrix}
 A_p-B_pD_r C_p& B_pC_r  \\ 
 -B_rC_p & A_r
\end{bmatrix}$,
$A_{R}=\begin{bmatrix}
 I_{n_p} & 0 \\ 
 0 & A_{R_r}
\end{bmatrix}$, $B_{cl}=\begin{bmatrix}
 B_pD_r & B_r \end{bmatrix}^T$, $C_{cl}=\begin{bmatrix}
C_p & 0 \end{bmatrix}$. 

\begin{example}
Modeling of a servomotor controlled by a fractional-order proportional-Clegg integrator (FPCI)
\end{example}

\begin{figure}[ptbh]
\begin{center}
\includegraphics[width=0.36\textwidth]{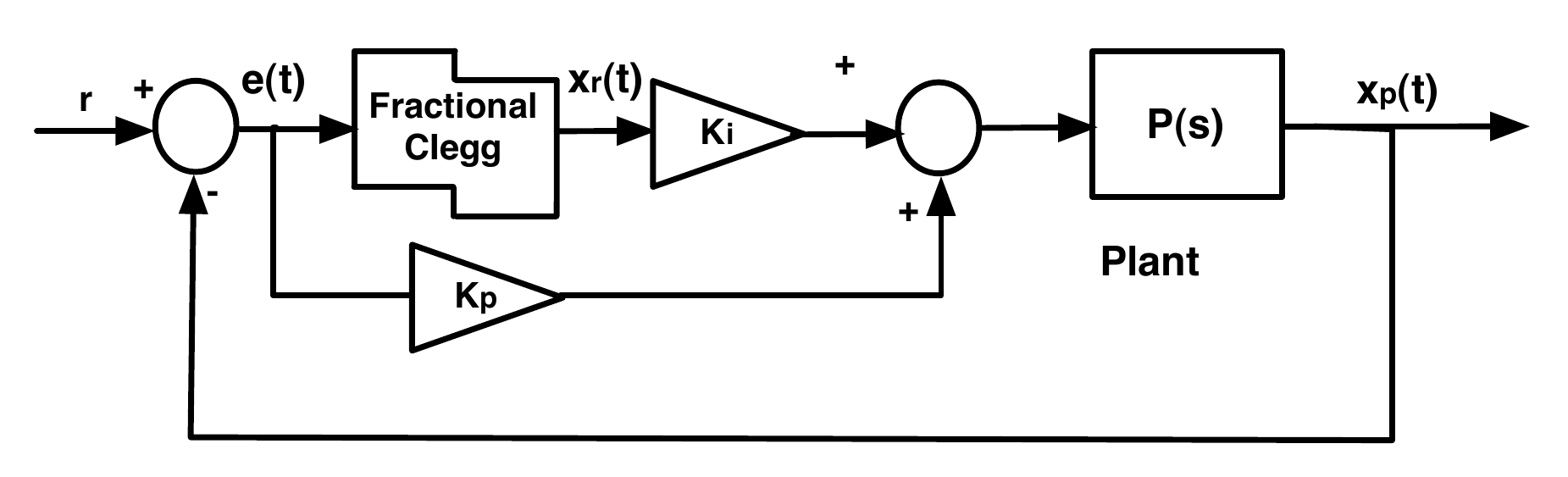}
\end{center}
\caption{Block diagram of a system controlled by a FPCI}
\label{Block_servo}%
\end{figure}

Consider the control scheme shown in see Fig.~\ref{Block_servo}, where the servomotor is given by
\begin{equation}
G_{s}(s)=\frac{K}{Ts+1}=\frac{0.93}{0.61s+1},
\label{servomodel}%
\end{equation}
and the FPCI by
\begin{equation}
R(s)=K_p+K_i\text{CI}^\alpha(s)=0.067+13.4\text{CI}^{0.75}(s),
\end{equation}
being CI$^\alpha$ a fractional Clegg integrator (FCI) (refer to part II \cite{HosseinNia2014b} for design details). Denote the state vector as $x(t)=(x_p(t), x_r(t))^T$, being $x_p(t)$ and $x_r(t)$ the plant and the controller states, respectively. Thus, the controlled system can be expressed of the form of (\ref{CLeq}) as follows:
\small
\begin{eqnarray}
\nonumber\begin{bmatrix}
\dot{x}_p(t)\\
D^\alpha x_r(t)
\end{bmatrix}=A_{cl}x(t)=\begin{bmatrix}
-\frac{1+KK_p}{\tau} &  \frac{KK_i}{\tau} \\
 -1 &  0 
\end{bmatrix}x(t)+\begin{bmatrix}
 \frac{KK_p}{\tau} \\
1
\end{bmatrix}r=\\
\begin{bmatrix}
-1.7415 &  20.4295 \\
 -1 &  0 
\end{bmatrix}x(t)+\begin{bmatrix}
 0.1021 \\
1
\end{bmatrix}r, \\
x(t^+)=A_Rx(t)=\begin{bmatrix}
1 & 0\\
 0 & 0 
\end{bmatrix}x(t),\ y(t)=C_{cl}x(t)=\begin{bmatrix}
1 & 0
\end{bmatrix}x(t).
\end{eqnarray}
\normalsize
Taking into account that $\alpha=0.75=\frac{3}{4}$, let consider $\mathcal{X}_{p_i}(t)=D^{\frac{i-1}{4}}x_p(t), i=1,\cdots, 4$ and $\mathcal{X}_{r_i}(t)=D^{\frac{i-1}{4}}x_r(t), i=1,\cdots, 3$ and define the state vector of the augmented system as $\mathcal{X}(t)=\left(\mathcal{X}_{p_1}(t),\cdots,\mathcal{X}_{p_4}(t),\mathcal{X}_{r_1}(t),\mathcal{X}_{r_2}(t),\mathcal{X}_{r_3}(t)\right)$, the augmented system can be represented as:
\begin{equation*}
D^{\frac{1}{4}} \mathcal{X}(t)=A\mathcal{X}(t)+Br=
\end{equation*}
\begin{equation}
\begin{bmatrix}
O_{3,1} & I_{3,3} & O_{3,1} & O_{3,2}  \\
-1.7415 & O_{1,3}  & 20.4295 & O_{1,2}  \\
 O_{2,1} & O_{2,3}  & O_{2,1} & I_{2,2}   \\
-1 & O_{1,3}  & 0 & O_{1,2}  
\end{bmatrix}\mathcal{X}(t)+\begin{bmatrix}
 O_{3,1}\\
 1.5246 \\
O_{2,1}\\
1
\end{bmatrix}r, 
\end{equation}
\begin{eqnarray}
\mathcal{X}(t^+)=\begin{bmatrix}
I_{6,6} & O_{6,1}\\
O_{1,6} & 0 
\end{bmatrix}\mathcal{X}(t),\ y(t)=\begin{bmatrix}
1 & O_{(1,6)} 
\end{bmatrix}\mathcal{X}(t),
\end{eqnarray}
where $O_{l,m}$ denotes a matrix of zeros with dimension of $l\times m$. 

\section{Stability of Fractional-Order Hybrid Systems}
\label{SFHSF}

Although stability of hybrid systems is typically analysed by Lyapunov's theory (see e.g.~\cite{Liberzon03,Narendra_94,Mori_98,Shim_98}), recently a frequency domain method equivalent to the common Lyapunov was proposed in \cite{Kunze08} to analyse the stability of a particular class of such systems. This section provides the stability conditions for two kinds of fractional-order hybrid systems, namely, switching and reset control systems, based on Lyapunov's theory and its frequency domain equivalence. Two examples of application are also given.

\subsection{Fractional-order switching systems}
\label{sec_fos}

The developed theory for fractional-order switching systems can be found in \cite{Hosseinnia2013b,Hosseinnia_12b,Hosseinnia13a}. Firstly, let us to recall the stability of fractional-order switching systems by common Lyapunov functions and its equivalence in frequency domain as preliminaries.

\begin{theorem}\text{(\cite{Hosseinnia2013b,Hosseinnia_12b})}
A fractional system described by (\ref{FSWHM}) with order $\alpha$, $1 \leq\alpha< 2$, is stable if and only if there exists a matrix $P=P^{T} > 0$, $P\in\mathbb{R}^{n \times n}$, such that
\small
\begin{align}%
\begin{bmatrix}
\left(  A_{i}^{T}P+PA_{i} \right) \sin \phi  &
\left(  A_{i}^{T}P-PA_{i} \right) \cos \phi  \\
\left(  -A_{i}^{T}P+PA_{i} \right) \cos \phi &
\left(  A_{i}^{T}P+PA_{i} \right) \sin \phi
\end{bmatrix}
<0,
\forall i=1,..., L,
\end{align}
\normalsize
where $\phi=\frac{\alpha\pi}{2}$.
\label{FSQ2}
\end{theorem}

\begin{theorem}\text{(\cite{Hosseinnia2013b,Hosseinnia_12b})}
A fractional system given by (\ref{FSWHM}) with order $\alpha$, $0<\alpha\leq1$, is stable if and only if there exists a matrix $P=P^{T} > 0$, $P
\in\mathbb{R}^{n \times n}$, such that
\begin{equation}
  \mathcal{A}_i^{T}P+P\mathcal{A}_i  <0,\ \ \forall i=1, ..., L.
\end{equation}
\label{FSQ00}
\end{theorem}

Next, frequency domain stability conditions will be given for fractional-order switching systems based on results in \cite{Kunze08}. Consider a stable pseudo-polynomial of order $n\alpha$ of system (\ref{FSWHM}) as
\begin{equation}
d(s)=s^{n\alpha}+d_{n-1}s^{(n-1)\alpha}+ \cdots+d_{1}s^{\alpha}+d_{0},
\end{equation}
and a polynomial of order $n$ of system $\dot{\tilde{x}}=\tilde{A}\tilde{x}$ as
\begin{equation}
c(s)=s^{n}+c_{n-1}s^{(n-1)}+ \cdots+c_{1}s +c_{0}. \label{TCP}%
\end{equation}

In the following, the necessary and sufficient condition for the
stability for fractional-order switching systems is given.

\begin{theorem} \text{(\cite{Hosseinnia2013b})}
Consider $d_{1}(s)$ and $d_{2}(s)$, two stable pseudo-polynomials of order
$n$ corresponding to the subsystems $D^{\alpha}x=A_{1}x$ and
$D^{\alpha}x=A_{2}x$ with order $\alpha$, $1\leq\alpha< 2$, respectively, then the following
statements are equivalent:

\begin{enumerate}
\item {%
$ \left|  \arg\left( \det((A_{1}^{2}-\omega^{2}I)-2j \omega A_{1}\sin \phi) \right) - \right. \\ \left. \arg\left(  \det((A_{2}^{2}-\omega^{2}I)-2j \omega A_{2}\sin \phi)\right) \right|  < \frac{\pi}{2}, \forall\omega
$,\\ being $I$ the identity matrix with proper dimensions.}

\item { $A_{1}$ and $A_{2}$ are stable, which means that $\exists P
=P^{T} >0 \in\mathbb{R}^{n\times n}$ such that
\begin{align*}%
\begin{bmatrix}
\left(  A_{i}^{T}P+PA_{i} \right) \sin \phi & \left(
A_{i}^{T}P-PA_{i} \right) \cos \phi\\
\left(  -A_{i}^{T}P+PA_{i} \right) \cos \phi & \left(
A_{i}^{T}P+PA_{i} \right) \sin \phi
\end{bmatrix}
<0,
\forall i=1, 2.
\end{align*}
}
\end{enumerate}

\label{Freq_stab_frac}
\end{theorem}

\begin{theorem} \text{(\cite{Hosseinnia2013b})}
Consider two stable fractional-order subsystems $D^{\alpha}x=A_{1}x$ and
$D^{\alpha}x=A_{2}x$ with order $\alpha$, $0<\alpha\leq1$, then the following
statements are equivalent:

\begin{enumerate}
\item {$\left|  \arg(\det(\mathcal{A}_{1} -j\omega I) )-\arg(\det
(\mathcal{A}_{2} -j\omega I))\right|  < \frac{\pi}{2}$, $\forall$ $\omega$.}

\item {$A_{1}$ and $A_{2}$ are stable, which means that $\exists P
=P^{T} >0 \in\mathbb{R}^{n\times n}$ such that }
\[
  \mathcal{A}_{i}   ^{T}P+P\mathcal{A}_{i}  <0, \forall i=1, 2.
\]
\end{enumerate}
 \label{Freq_stab_frac0}
\end{theorem}

Although the theory developed in the frequency domain does not necessarily prove the strictly positive realness, a relation equivalent to the stability was obtained. See \cite{HosseinNia2013t} for the switching systems more than two subsystems.

\begin{example}
\label{expg}
Stability of a fractional-order switching system with two subsystems
\end{example}

\begin{figure}[ptbh]
\begin{center}
\includegraphics[width=0.37\textwidth]{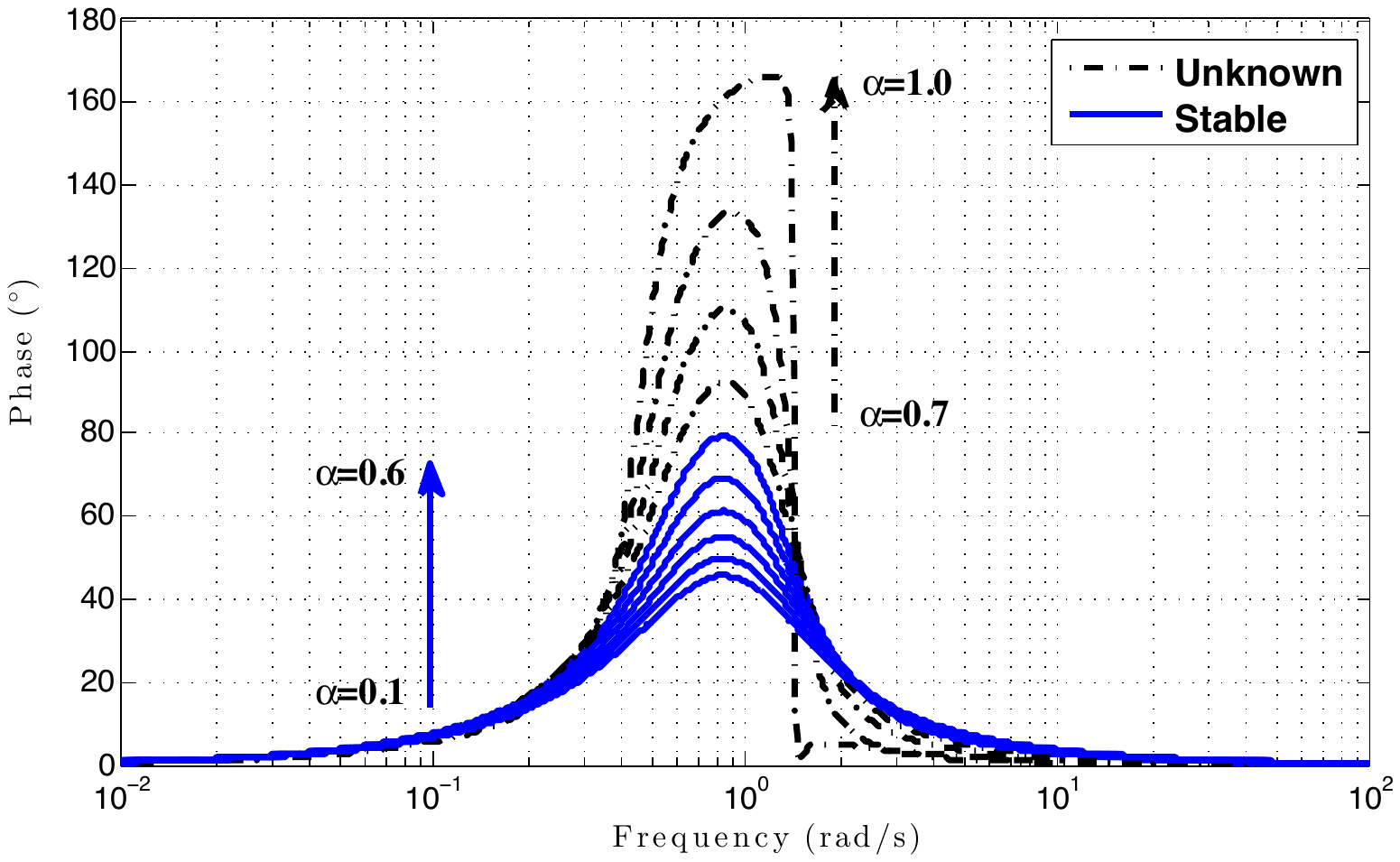}
\end{center}
\caption{Phase differences of characteristic polynomials of system in Example~\ref{expg} for different values of its order $\alpha$, $0<\alpha \leq 1$}%
\label{Expgsw}%
\end{figure}

Consider the switching system (\ref{FSWHM}) with $L=2$ with the following parameters:
$A_{1}=%
\begin{bmatrix}
 -0.1  &  0.1\\
   -2.0  &  -0.1
\end{bmatrix}
$, $A_{2}=%
\begin{bmatrix}
 -0.01  &  2.0\\
   -0.1  &  -0.01
\end{bmatrix}$ and order $\alpha$, $0<\alpha \leq 1$. Applying Theorem~\ref{Freq_stab_frac0}, the phase difference condition should be satisfied for all $\alpha$, $0<\alpha \leq 1$, to guarantee the stability --this condition is depicted in Fig.~\ref{Expgsw} for $0<\alpha \leq 1$ with increments of $0.1$. As can be seen, the fractional-order system is stable for $\alpha\in(0,0.6]$. The phase differences when $\alpha\in\lbrack0.7,1]$ are bigger than ${\pi}/{2}$ which indicates unknown stability status, i.e., the system may be stable or unstable. For more details see \cite{Hosseinnia2013b}.

\begin{example}
\label{ex:SW}
Stability analysis of the SmartWheel controlled by fractional-order gain scheduled controller
\end{example}
In the literature, it is widely noticed that systems controlled through networks exhibit high switching behaviours and thus their design and analysis within the switching system framework are highly desirable (refer to e.g. \cite{Wang08,Branicky2005,Hristu05}). The case of study to be considered is the Internet-based control of a platform, called SmartWheel, placed at the Center for Self-Organizing and Intelligent Systems (CSOIS), Utah State University, USA, from the University of Extremadura, Spain. Thus, the existence of network time-varying delays together with the application of gain scheduling result in the transformation of
the closed-loop system into a switching system with finite number of subsystems as follows (the full description can be found in \cite{Tejado13a,Tejado2011}):
\begin{equation}
G_{j}(s)=\frac{0.1484}{0.045s+1}e^{-(0.592+\tau_{j})s},\label{Sys_Cont}
\end{equation}
\begin{equation}
C_{j}(s)=\beta_{j}\left(  2.1586+\frac{5.9853}{s^{1.1}}\right)  ,\text{
}j=1,2,...,13,\label{Sys_Cont2}
\end{equation}
where $\tau_j$ refers to the network delay $\tau_{network}$ and $\beta_j$ is the gain scheduler with the switching parameters given in Table~\ref{TIDC}. Hence, there are $13$ subsystems to be considered.

\begin{figure}[htbp]
\begin{center}
\includegraphics[width=0.37\textwidth]{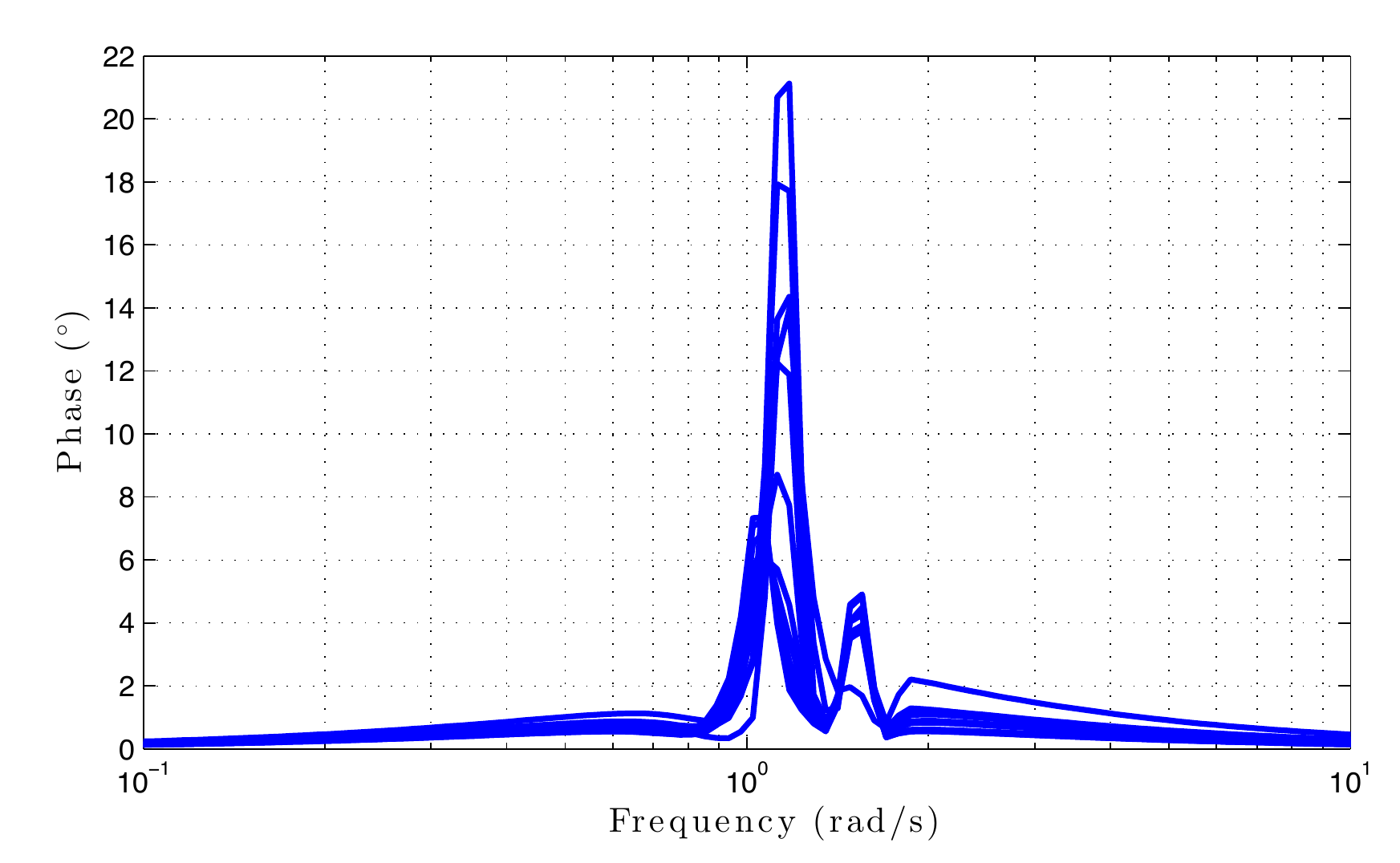}\\
\caption{Phase differences of characteristic polynomials of system in Example~\ref{ex:SW} given by conditions (\ref{stabeq1})--(\ref{stabeq3})}
\label{SmWhStability}
\end{center}
\end{figure}

In order to apply Theorem \ref{Freq_stab_frac0} the controlled system has to be described in the form of commensurate-order system. Therefore, assuming Pad\'e approximation of delay is
\begin{equation*}
\text{Pade}(e^{-(0.592+\tau_{j})s})=\frac{P_n(s)_j}{P_d(s)_j},
\end{equation*}
the closed-loop pseudo characteristic polynomials can be represented as follows:
\begin{equation}
d_j(s)=P_d(s)\left(s^{2.1}+22.22s^{1.1}\right)+\beta_jP_n(s)\left(7.12s+19.74\right).
\end{equation}
Defining $\lambda=s^{0.1}$, the characteristic polynomials of the system can be obtained as
\small
\begin{eqnarray}
\nonumber c_j(\lambda^{10})=P_d(\lambda^{10})\left(\lambda^{21}+22.22\lambda^{11}\right)+\\
\beta_jP_n(\lambda^{10})\left(7.12\lambda^{10}+19.74\right).
\end{eqnarray}
\normalsize
Suppose $c_j(\lambda)=\bf{c}$$_j\begin{bmatrix} \lambda^{21+m} & \lambda^{m+20} & \cdots & 1 \end{bmatrix}$, where $\bf{c}$$_j=\begin{bmatrix} 1 & \bf{c}_{\it{m}+\text{20}}^{\it{j}} & \cdots & \bf{c}_{\text{0}}^{\it{j}} \end{bmatrix}$ is a vector with $m+22$ elements and $m$ is order of Pad\'e approximation. Hence, the commensurate fractional-order system can be realised as
\begin{equation}
D^{0.1}x=A_jx=\begin{bmatrix} 
-\bf{c}_{\it{m}+\text{20}}^{\it{j}} & \cdots &  -\bf{c}_{\text{1}}^{\it{j}} & -\bf{c}_{\text{0}}^{\it{j}}   \\
1 & \cdots & 0 & 0  \\
\vdots & \ddots & \vdots &\vdots \\
0 & \cdots & 1 & 0  
\end{bmatrix}x, x\in\mathbb{R}^{m+21}.
\end{equation}
Now, we can easily apply Theorem \ref{Freq_stab_frac0} to analyse the stability of the system. The following $12$ conditions should be satisfied to guarantee the stability of the controlled system:
\begin{eqnarray}
\nonumber \left|  \arg\left( \det((A_{1}^{2}-\omega^{2}I)-2j \omega A_{1}\sin \phi) \right) - \right. \\ \left. \arg\left(  \det((A_{2}^{2}-\omega^{2}I)-2j \omega A_{2}\sin \phi)\right) \right|  < \frac{\pi}{2}, \forall\omega,
 \label{stabeq1}
\end{eqnarray}
\begin{eqnarray}
\nonumber \left|  \arg\left( \det((A_{2}^{2}-\omega^{2}I)-2j \omega A_{2}\sin \phi) \right) - \right. \\ \left. \arg\left(  \det((A_{3}^{2}-\omega^{2}I)-2j \omega A_{3}\sin \phi)\right) \right|  < \frac{\pi}{2}, \forall\omega,
 \label{stabeq2}
\end{eqnarray}
\begin{eqnarray*}
\vdots
\end{eqnarray*}
\begin{eqnarray}
\nonumber \left|  \arg\left( \det((A_{j-1}^{2}-\omega^{2}I)-2j \omega A_{j-1}\sin \phi) \right) - \right. \\ \left. \arg\left(  \det((A_{j}^{2}-\omega^{2}I)-2j \omega A_{j}\sin \phi)\right) \right|  < \frac{\pi}{2}, \forall\omega,
 \label{stabeq3}
\end{eqnarray}
where $\phi=\frac{\alpha\pi}{2}$. The simulation of conditions (\ref{stabeq1})--(\ref{stabeq3}) is shown in Fig. \ref{SmWhStability}. It can be observed that the maximum phase difference is less than $90^\circ$ and, consequently, the system is stable.

\begin{center}
\begin{table*}[ptbh]
\caption{System and controller parameters in Example \ref{ex:SW} for each switching}
\begin{center}
{\footnotesize
\begin{tabular}
[c]{|l|c|c|c|c|c|c|c|c|c|c|c|c|c|}\hline
$j$ & $1$ & $2$ & $3$ & $4$ & $5$ & $6$ & $7$ & $8$ & $9$ & $10$ & $11$ & $12$ & $13$\\\hline
$\tau$ & $0$ & $0.1$ & $0.2$ & $0.3$ & $0.4$ & $0.5$ & $0.6$ & $0.7$ & $0.8$ & $0.9$ & $1.0$ & $1.1$ & $1.2$\\\hline
$\beta$ & $1.6$ & $1.35$ & $1.3$ & $1.15$ & $1$ & $0.9$ & $0.8$ & $0.7$ & $0.65$ & $0.6$ & $0.55$ & $0.5$ & $0.45$\\\hline
\end{tabular}
\label{TIDC}%
}
\end{center}
\end{table*}
\end{center}

\subsection{Fractional-order reset control systems}
\label{sec_reset}

In this section, stability of fractional-order reset control systems is analysed using the Lyapunov-like method presented previously. This theory was proposed in \cite{Hosseinnia13b}.

\begin{definition} Reset control system (\ref{CLeq}) is said to satisfy the H$_\beta$-condition if there exists a $\beta \in \mathbb{R}^{n_{\mathcal{R}}}$ and a positive-definite matrix $P_{\mathcal{R}} \in \mathbb{R}^{n_{\mathcal{R}} \times n_{\mathcal{R}}}$ such that
\begin{eqnarray}
H_\beta(s)=\begin{bmatrix}
\beta C_p & 0_{n_{\bar{\mathcal{R}}}} & P_{\mathcal{R}}
\end{bmatrix}\left ( sI-\mathcal{A} \right )^{-1}\begin{bmatrix}
0\\ 
0^T_{\bar{\mathcal{R}}}\\ 
I_{\mathcal{R}}
\end{bmatrix},
\label{Hb}
\end{eqnarray}
where $\mathcal{A}= \left( -\left(-A_{cl}\right)  ^{\frac{1}{2-\alpha}}\right)$.
\end{definition} 

In accordance with \cite{ioannou1987}, it is obvious that the H$_\beta(s)$ is strictly positive real (SPR) if 
\begin{eqnarray}
\label{FrSPR}
\left |\arg(H_\beta(j\omega)) \right |<\frac{\pi}{2}, \forall \omega.
\end{eqnarray} 

\begin{theorem}\text{(\cite{Hosseinnia13b})} The closed-loop fractional-order reset control system (\ref{CLeq}) is asymptotically stable if and only if it satisfies the H$_\beta$-condition (\ref{Hb}) or its phase equivalence (\ref{FrSPR}).
\label{FRQS}
\end{theorem}

An example of application is given next.

\begin{example}
\label{Rest_exp_stab}
Stability of a fractional-order reset control system
\end{example}

\begin{figure}[ptbh]
\begin{center}
\includegraphics[width=0.37\textwidth]{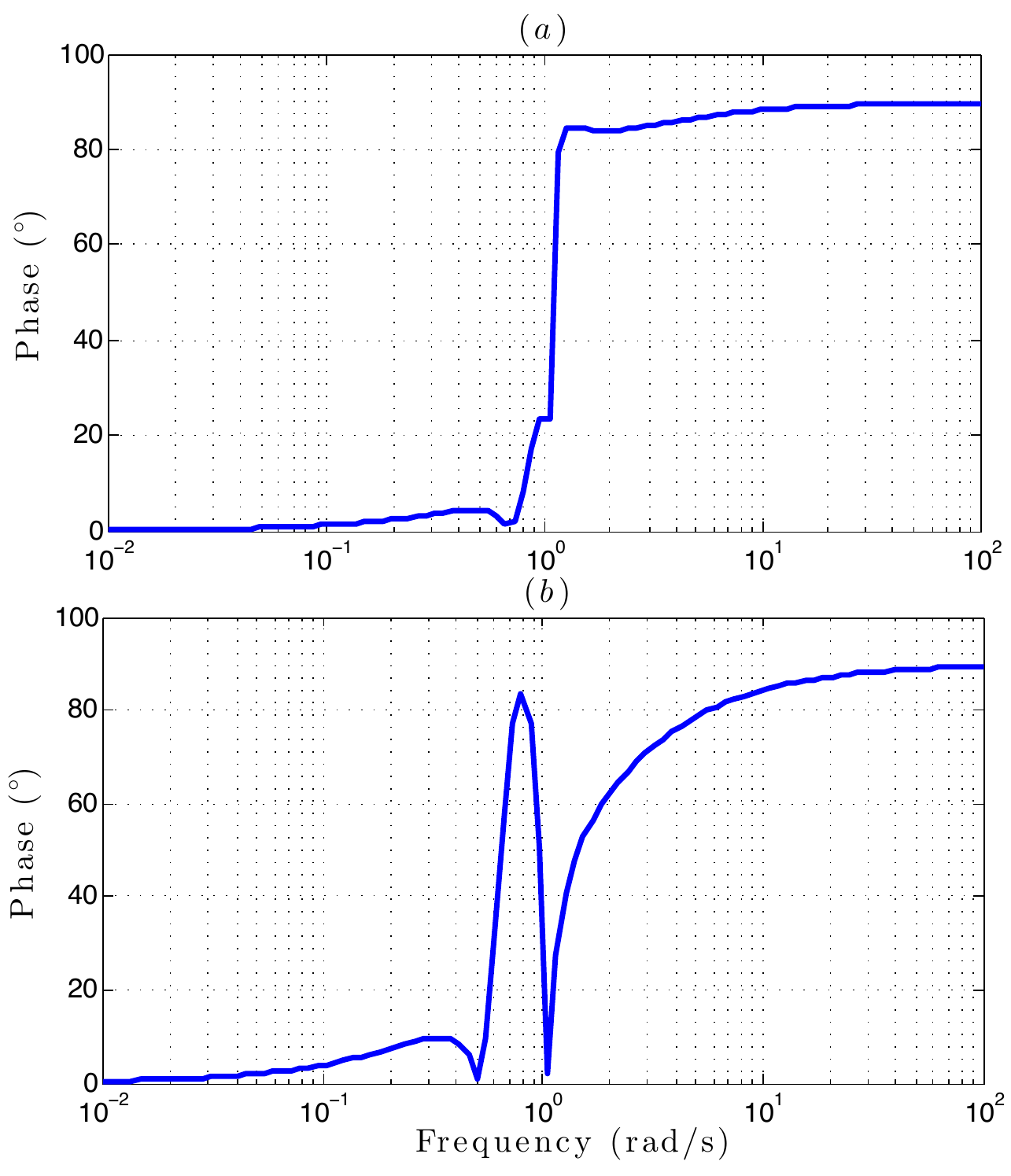}
\end{center}
\caption{Phase equivalence of H$_\beta$ in Example \ref{Rest_exp_stab}: ($a$) Applying FCI ($b$) Applying FORE}
\label{HB_FCIexp2}%
\end{figure}

Let us consider the same feedback system as in \cite{hollot2001} with the following system, base controller and reset controller transfer functions: $P(s)=\frac{1}{s^2+0.2s}$, $C(s)=s+1$ and $R(s)=\frac{1}{s^\alpha+b}$,
respectively. The system stability will be analysed for different reset controllers: the first-order reset element (FORE) controller, with $b\neq0$ and $\alpha=1$, the CI, with $b=0$ and $\alpha=1$, and the FCI, with $b=0$ and $\alpha=0.5$. For FORE controller, the integer-order closed-loop system can be given by:
\small
\begin{eqnarray*}
\left\{
\begin{matrix}
\dot{x}(t)=A_{cl}x=\begin{bmatrix}
0 & 1 & 0\\
0 & -0.2 & 1\\
 -1 & -1 & -b 
\end{bmatrix}x(t)\\ 
x(t^+)=A_Rx=\begin{bmatrix}
1 & 0 & 0\\
0 & 1 & 0\\
 0 & 0 & 0 
\end{bmatrix}x(t)\\ 
y=C_{cl}x=\begin{bmatrix}
1 & 1 & 0
\end{bmatrix}x(t)
\end{matrix}
\right.
\end{eqnarray*}
\normalsize
where $x(t)=\left[x_{p_1}(t),x_{p_2}(t),x_r(t)\right]^T$. And, the closed-loop system using FCI can be stated as
\small
\begin{eqnarray*}
\left\{
\begin{matrix}
D^{0.5}\mathcal{X}(t)=\mathbf{A} _{cl}\mathcal{X}(t)=\begin{bmatrix}
0 & 1 & 0 & 0 & 0\\
0 & 0 & 1 & 0 & 0\\
0 & 0 & 0 & 1 & 0\\
0 & 0 & -0.2 & 0 & 1\\
 -1 & 0 & -1 & 0 & 0
\end{bmatrix}\mathcal{X}(t)\\ 
\mathcal{X}(t^+)=\mathbf{A}_R\mathcal{X}(t)=\begin{bmatrix}
I_4 & 0_{4,1} \\
 0_{1,4} & 0 
\end{bmatrix}\mathcal{X}(t)\\ 
y=\mathbf{C}_{cl}\mathcal{X}(t)=\begin{bmatrix}
 1 & 0 & 1 & 0 & 0
\end{bmatrix}\mathcal{X}(t)
\end{matrix}
\right.
\end{eqnarray*}
\normalsize
where $\mathcal{X}(t)=\left[\mathcal{X}_{p_1}(t),\cdots,\mathcal{X}_{p_4}(t),x_r(t)\right]^T$, $\mathcal{X}_{p_1}(t)=x_{p_1}(t)$, $\mathcal{X}_{p_3}(t)=x_{p_2}(t)$. According to (\ref{Hb}), H$_\beta$-conditions corresponding to FORE and FCI controllers are, respectively, given by (for both cases, $n_{\mathcal{R}}=1$ and, then, $P_{\mathcal{R}}=1$):
\small
\begin{equation*}
\nonumber
H^{FORE}_\beta(s)= \begin{bmatrix}
\beta & 0 & 1
\end{bmatrix}\left ( sI-A_{cl} \right )^{-1}\begin{bmatrix}
0\\ 
0\\ 
1
\end{bmatrix}=
\end{equation*}
\begin{equation}
\frac{s^2+0.2s+0.8\beta}{s^3+(b+0.2)s^2+(1+0.2b)s+1},
\label{HB_FORE_CI}
\end{equation}
\normalsize
\small
\begin{equation}
H^{FCI}_\beta(s)=
\begin{bmatrix}
\beta & 0 & \beta & 0 & 1
\end{bmatrix}\left ( sI-\left( -\left(-\mathbf{A}_{cl}\right)  ^{\frac{2}{3}}\right) \right )^{-1}\begin{bmatrix}
0\\ 
0\\ 
0\\ 
0\\ 
1
\end{bmatrix}.
\label{HB_FCI}
\end{equation}
\normalsize
Using Theorem \ref{FRQS}, the closed-loop systems controlled by FORE and FCI are asymptotically stable if H$^{FORE}_\beta(s)$ and H$^{FCI}_\beta(s)$ are SPR. Substituting $b=1$ in (\ref{HB_FORE_CI}), the FORE reset system is asymptotically stable for all $0.42<\beta \leq1.46$. With respect to CI (similarly to FORE but with $b=0$), stability cannot be guaranteed with this theorem. And applying FCI, it can be easily stated that the system is asymptotically stable for $\beta \leq 0.62$. In addition, the phase equivalences corresponding to (\ref{HB_FORE_CI}) and (\ref{HB_FCI}) are shown in Fig. \ref{HB_FCIexp2} for $\beta=0.5$ and $b=1$. It can be seen that both phases verifies condition (\ref{FrSPR}), which has concordance with the theoretical results.

\section{Conclusions}
\label{Concf}

In part I of this paper, modeling of fractional-order hybrid systems (FHS) was introduced based on fractional-order differential inclusions, especially for two special kinds of them, i.e., switching and reset control systems. Moreover, stability of such FHS was also analysed based on Lyapunov's theory and its frequency domain equivalence. Some examples were given to show the way of modeling and the applicability of the developed stability theory. 

Since there is no a general agreement of the interpretation of state space representation of fractional-order systems, mainly concerning initial values (see e.g.~\cite{Sabatier13}), a further study should be carried out for fractional-order reset control taking into account this issue in future work.

\bibliographystyle{IEEEtran}

\begin{thebibliography}{10}
\providecommand{\url}[1]{#1}
\csname url@samestyle\endcsname
\providecommand{\newblock}{\relax}
\providecommand{\bibinfo}[2]{#2}
\providecommand{\BIBentrySTDinterwordspacing}{\spaceskip=0pt\relax}
\providecommand{\BIBentryALTinterwordstretchfactor}{4}
\providecommand{\BIBentryALTinterwordspacing}{\spaceskip=\fontdimen2\font plus
\BIBentryALTinterwordstretchfactor\fontdimen3\font minus
  \fontdimen4\font\relax}
\providecommand{\BIBforeignlanguage}[2]{{%
\expandafter\ifx\csname l@#1\endcsname\relax
\typeout{** WARNING: IEEEtran.bst: No hyphenation pattern has been}%
\typeout{** loaded for the language `#1'. Using the pattern for}%
\typeout{** the default language instead.}%
\else
\language=\csname l@#1\endcsname
\fi
#2}}
\providecommand{\BIBdecl}{\relax}
\BIBdecl

\bibitem{Gollu_89}
A.~Gollu and P.~Varaiya, ``Hybrid dynamical systems,'' in \emph{Proceedings of
  the 28th IEEE Conference on Decision and Control}.\hskip 1em plus 0.5em minus
  0.4em\relax IEEE, 1989, pp. 2708--2712.

\bibitem{schumacher_99}
A.~J. van~der Schaft and J.~M. Schumacher, \emph{Introduction to hybrid
  dynamical systems}.\hskip 1em plus 0.5em minus 0.4em\relax Springer-Verlag,
  1999.

\bibitem{Goebel_09}
R.~Goebel, R.~Sanfelice, and A.~Teel, ``Hybrid dynamical systems,''
  \emph{Control Systems Magazine, IEEE}, vol.~29, no.~2, pp. 28--93, 2009.

\bibitem{banos2011}
A.~Ba\~nos and A.~Barreiro, \emph{Reset Control Systems}.\hskip 1em plus 0.5em
  minus 0.4em\relax Springer Verlag, 2011.

\bibitem{Schutter2009}
B.~D. Schutter, W.~Heemels, J.~Lunze, and C.~Prieur, \emph{Handbook of Hybrid
  Systems Control--Theory, Tools, Applications}.\hskip 1em plus 0.5em minus
  0.4em\relax Cambridge University Press, 2009, pp. 31--35.

\bibitem{Monje10}
C.~A. Monje, Y.~Q. Chen, B.~M. Vinagre, D.~Xue, and V.~Feliu,
  \emph{Fractional-order Systems and Controls. Fundamentals and
  Applications}.\hskip 1em plus 0.5em minus 0.4em\relax Springer, 2010.

\bibitem{Podlubny_99a}
I.~Podlubny, \emph{Fractional Differential Equations. An Introduction to
  Fractional Derivatives, Fractional Differential Equations, Some Methods of
  Their Solution and Some of Their Applications}.\hskip 1em plus 0.5em minus
  0.4em\relax Academic Press, San Diego - New York - London, 1999.

\bibitem{Sun_05}
Z.~Sun and S.~S. Ge, \emph{Switched Linear Systems: Control and Design}.\hskip
  1em plus 0.5em minus 0.4em\relax Springer-Verlag, 2005.

\bibitem{HosseinNia2014b}
S.~H. HosseinNia, I.~Tejado, and B.~M. Vinagre, ``Hybrid systems and control
  with fractional dynamics ({II}): {C}ontrol,'' in \emph{Proceedings of the
  2014 International Conference in Fractional Differentiation and its
  Applications (ICFDA'14)}, 2014.

\bibitem{Liberzon03}
D.~Liberzon, \emph{Switching in Systems and Control}.\hskip 1em plus 0.5em
  minus 0.4em\relax Birk\"{a}user, 2003.

\bibitem{Narendra_94}
K.~S. Narendra and J.~Balakrishnan, ``A common {L}yapunov function for stable
  {LTI} system with commuting {A}-matrices,'' \emph{IEEE Transactions on
  Automatic Control}, vol.~39, no.~12, pp. 2469--2471, 1994.

\bibitem{Mori_98}
Y.~Mori, T.~Mori, and Y.~Kuroe, ``{O}n a class of linear constant systems which
  have a common quadratic lyapunov function,'' in \emph{Proceedings of the 37th
  IEEE Conference on Decision and Control}, 1998.

\bibitem{Shim_98}
H.~Shim, D.~Noh, , and J.~Seo, ``Common lyapunov function for exponentially
  stable nonlinear systems,'' in \emph{Proceedings of the 4th SIAM Conference
  on Control and its Applications}, 1998.

\bibitem{Kunze08}
M.~Kunze, A.~Karimi, and R.~Longchamp, ``Frequency domain controller design by
  linear programming guaranteeing quadratic stability,'' in \emph{Proceedings
  of the 47th Conference on Decision and Control (CDC'08)}, 2008, pp. 345--350.

\bibitem{Hosseinnia2013b}
S.~H. HosseinNia, I.~Tejado, and B.~M. Vinagre, ``Stability of fractional order
  switching systems,'' \emph{Computer \& Mathematics with Applications},
  vol.~66, no.~5, pp. 585--596, 2013.

\bibitem{Hosseinnia_12b}
------, ``Stability of fractional order switching systems,'' in
  \emph{Proceedings of the 5th Workshop on Fractional Differentiation and Its
  Applications (FDA'12)}, 2012.

\bibitem{Hosseinnia13a}
------, ``Basic properties and stability of fractional order reset control
  systems,'' in \emph{Proceedings of the 12th European Control Conference
  (ECC'13)}, 2013.

\bibitem{HosseinNia2013t}
S.~H. HosseinNia, ``Fractional hybrid control systems: Modeling, analysis and
  applications to mobile robotics and mechatronics,'' Ph.D. dissertation,
  University of Extremadura, 2013.

\bibitem{Wang08}
F.-Y. Wang and D.~Liu, Eds., \emph{Networked Control Systems: Theory and
  Applications}.\hskip 1em plus 0.5em minus 0.4em\relax Springer-Verlag, 2008.

\bibitem{Branicky2005}
M.~S. Branicky, ``Introduction to hybrid systems,'' in \emph{Handbook of
  Networked and Embedded Control Systems}, D.~Hristu-Varsakelis and W.~Levine,
  Eds.\hskip 1em plus 0.5em minus 0.4em\relax Birkh\"{a}user Boston, 2005, pp.
  91--116.

\bibitem{Hristu05}
R.~Alur, K.-E. Arzen, J.~Baillieul, T.~Henzinger, D.~Hristu-Varsakelis, and
  W.~S. Levine, \emph{Handbook of networked and embedded control
  systems}.\hskip 1em plus 0.5em minus 0.4em\relax Birkh{\"a}user Boston, 2005.

\bibitem{Tejado13a}
I.~Tejado, S.~H. HosseinNia, B.~M. Vinagre, and Y.~Q. Chen, ``Efficient control
  of a {S}mart{W}heel via internet with compensation of variable delays,''
  \emph{Mechatronics}, vol.~23, pp. 821--827, 2013.

\bibitem{Tejado2011}
I.~Tejado, ``Some contributions in networked control systems based on
  fractional calculus,'' Ph.D. dissertation, University of Extremadura, Spain,
  2011.

\bibitem{Hosseinnia13b}
S.~H. HosseinNia, I.~Tejado, and B.~M. Vinagre, ``Fractional-order reset
  control: {A}pplication to a servomotor,'' \emph{Mechatronics}, vol.~23,
  no.~7, pp. 781--788, 2013.

\bibitem{ioannou1987}
P.~Ioannou and G.~Tao, ``Frequency domain conditions for strictly positive real
  functions,'' \emph{IEEE Transactions on Automatic Control}, vol.~32, no.~1,
  pp. 53--54, 1987.

\bibitem{hollot2001}
C.~Hollot, O.~Beker, Y.~Chait, and Q.~Chen, ``On stablishing classic
  performance measures for reset control systems,'' \emph{Perspectives in
  robust control}, pp. 123--147, 2001.

\bibitem{Sabatier13}
J.~Sabatier, C.~Farges, and J.-C. Trigeassou, ``Fractional systems state space
  description: Some wrong ideas and proposed solutions,'' \emph{Journal of
  Vibration and Control}, 2013.

\end{thebibliography}

\end{document}